\documentclass[showpacs,twocolumn,prd,nofootinbib]{revtex4-1}
 \usepackage{color,graphicx,epsfig}
 \usepackage{amsmath,amssymb,dsfont,epsfig,graphicx,xcolor}
 \usepackage{ifpdf}
 \usepackage{amsmath}
 \usepackage{bm}
 \usepackage{color}
 \usepackage[english]{babel}
 \usepackage{graphicx}%
 \usepackage{amsfonts}%
 \usepackage{amssymb}
 \usepackage{braket}
 \usepackage{hyperref}
 \usepackage{epstopdf}
 \usepackage{slashed}
\usepackage{float}
 \graphicspath{{Plots/}}
 \usepackage{commath}

\begin{document}

\def\LjubljanaFMF{Faculty of Mathematics and Physics, University of Ljubljana,
 Jadranska 19, 1000 Ljubljana, Slovenia }
\def\LjubljanaIJS{Jo\v zef Stefan Institute, Jamova 39, 1000 Ljubljana, Slovenia}

\title{Coloured Scalars Mediated Rare Charm Meson Decays to Invisible Fermions}

\author{Svjetlana Fajfer}
\email[Electronic address:]{svjetlana.fajfer@ijs.si} 
\affiliation{\LjubljanaIJS}
\affiliation{\LjubljanaFMF}
\author{Anja Novosel}
\email[Electronic address:]{anja.novosel@ijs.si} 
\affiliation{\LjubljanaIJS}
\affiliation{\LjubljanaFMF} 

\date{\today}
\begin{abstract}
We consider effects of coloured scalar mediators in  decays $c\to u$ {\it invisibles}.  In particular, in these processes, as invisibles, we consider massive right-handed fermions. 
The coloured scalar $\bar S_1\equiv (\bar 3, 1, -2/3)$, due to its coupling to weak singlets up-quarks and  invisible right-handed fermions ($\chi$), is particularly interesting. Then, we consider $\tilde R_2 \equiv (\bar 3, 2, 1/6)$, which as  a weak doublet is a subject of severe low-energy constraints. The $\chi$  mass is considered in the range $(m_K -m_\pi)/2\leq m_\chi \leq (m_D - m_\pi)/2$.  We determine branching ratios for $D\to \chi \bar \chi$, $D\to \chi \bar \chi \gamma$ and $D\to \pi \chi \chi$ for several $\chi$ masses, using most constraining bounds. For  $\bar S_1$, the  most constraining is $D^0 -\bar D^0$ mixing, while in the case of $\tilde R_2$ the strongest constraint comes from $B\to K \slashed{E}$. We find  in decays mediated by $\bar S_1$  that branching ratios can be  $\mathcal B(D\to \chi \bar \chi)< 10^{-8}$  for $m_\chi=0.8$ GeV,  $\mathcal B(D\to \chi \bar \chi \gamma) \sim 10^{-8}$  for $m_\chi=0.18$ GeV, while $\mathcal B(D^+ \to \pi^+ \chi \bar \chi )$ can reach $ \sim 10^{-8}$ for $m_\chi=0.18$  GeV. In the case of $\tilde R_2$ these decay rates are very suppressed. We find that future tau-charm factories and Belle II experiments offer good opportunities to search for such processes.  Both $\bar S_1$ and $\tilde R_2$ might have masses within LHC reach.

\end{abstract}

\maketitle
\section{Introduction}
Low-energy constraints  of physics beyond  Standard
Model (BSM) are well established for down-like quarks 
by numerous searches in processes with hadrons containing one $b$ or/and $s$ quark.   However, in the up-quark sector, searches are performed in top decays, suitable
for LHC studies, while in charm hadron processes  at b-factories or/and $\tau$-charm factories. 
Recently, an extensive study on $c\to u \nu \bar \nu$ appeared in Ref.  \cite{Bause:2020xzj}, 
pointing out that observables very small in the Standard
Model (SM) offer unique (null) tests of BSM physics. Namely, for  charm flavour changing neutral current (FCNC) 
 processes, severe Glashow-Iliopoulos-Maiani (GIM) 
suppression occurs. The decay $D^0 \to \nu \bar \nu$ amplitude is helicity suppressed
in the SM. The authors of \cite{Badin:2010uh} made very detailed study
of heavy meson decays to invisibles, assuming that the
invisibles can be scalars or fermions with both helicites. They found out that in the SM branching ratio 
 ${\mathcal B}(D^0 \to \nu \bar \nu ) = 1.1 \times 10^{-31}$. Then the authors of \cite{Bhattacharya:2018msv}  found
that the decay width of $D^0 \to {\it invisibles}$ in the SM
is actually dominated by the contribution of $D^0 \to \nu \bar \nu \nu \bar \nu$.
These studies'  main message is that SM provides
no irreducible background to analysis of invisibles in decays of charm (and beauty) mesons. They also suggested \cite{Badin:2010uh}, 
that in searches for a Dark Matter candidate, it might
be important to investigate process with $\chi \bar \chi \gamma$ in the final
state, since a massless photon eliminates the helicity
suppression. We also determine branching ratios for such decay modes. The authors of Ref. \cite{Bause:2020xzj} computed the expected event rate for the charm hadron decays to a final hadronic state and neutrino - anti-neutrino
states. They found out that in experiments like Belle II,  which can reach per-mile efficiencies or better, these processes can be seen.  In addition  future FCC-ee might
be capable of measuring branching ratios of  $\mathcal{O}(10^{-6})$ down to $\mathcal{O}(10^{-8})$, in particular $D^0$, $D^+_{(s)}$ and $\Lambda_c^+$ decay modes.

On the other hand, the  Belle collaboration  already reached bound of the branching ratio for 
${\mathcal B}(D^0\to {\it invisibles}) = 9.4\times 10^{-5}$  and the Belle II experiment is expected to improve it \cite{Kou:2018nap}. 
The other $e^+ e^-$ machines as BESSIII \cite{Ablikim:2019hff} and future FCC-ee  running colliders at the $Z$ energies \cite{Abada:2019lih,Abada:2019zxq} with a significant charm production with ${\mathcal B} (Z\to c \bar c)\simeq 0.22$ \cite{Abada:2019zxq} provide  us with excellent tools for precision study of charm decays.
 
  In this work we focus on the particular scenarios with coloured scalars or leptoquarks as mediators of the invisible fermions interaction with quarks. The coloured scalar might have the electric charge of $2/3$ or $-1/3$ depending on the interactions with up or down quarks. Instead of using general assumption on the lepton flavour structure from \cite{Bause:2020xzj}  and justifying  Belle  bound from \cite{Lai:2016uvj},  we rely on observables coming from the $D^0 - \bar D^0$  oscillations and in the case of weak doublets, we include constraints from other flavour processes. 

Motivated by previous works of Refs. \cite{Bause:2020xzj,Bause:2020obd,Badin:2010uh,Golowich:2009ii,MartinCamalich:2020dfe,Faisel:2020php}, we investigate $c \to u \bar \chi \chi$ with $\chi$ being a massive $SU(2)_L$ singlet. 
Coloured scalars carry out interactions between invisible fermions and quarks.  Namely, leptoquarks usually denote the boson interacting with quarks and leptons. However, the state $\bar S_1$ does not interact with the SM leptons and, therefore,  it is more appropriate to call it coloured scalar. Our approach is rather minimalistic due to only two Yukawa couplings and the mass of coloured scalar. The effective Lagrangian and coloured scalar mediators are introduced in Sec. II. 
In Sec. III we describe effects of $\bar S_1$ mediator in rare charm decays, while in Sec. IV we give details of $\tilde R_2$ mediation in the same processes.
Sec. V contains conclusions and outlook.
\section{ Coloured Scalars in $c\to u \chi \bar \chi$}
In experimental searches, the transition $c\to u \,{\it invisibles}$ might be approached in processes $c\to u \slashed{E}$ with $ \slashed{E}$ being missing energy. Therefore, invisibles can be either SM neutrinos or new right-handed neutral fermions (having quantum numbers of right-handed neutrinos), or scalars/vectors as suggested in Ref. \cite{Badin:2010uh}.  
The authors of Refs. \cite{Bause:2020xzj,Bause:2020obd}  considered in detail general framework of New Physics (NP) in $c \to u$ {\it invisibles}, relying on $SU(2)_L$ invariance and data on charged lepton processes \cite{Bause:2020obd}. They found that these assumptions allow upper limits as large as few $10^{-5}$, while in the limit of lepton universality branching ratios can be as large as $10^{-6}$. To consider invisible fermions, having quantum numbers of right-handed neutrinos, and being massive, we extend the effective Lagrangian by additional operators as described in Refs. \cite{Bhattacharya:2018msv,Dorsner:2016wpm}


\begin{align} \label{LQeffL}
\begin{aligned}
\mathcal{L}_{\text{eff}}=&\sqrt{2}G_F\bigg[c^{LL}(\overline{u}_L\gamma_\mu c_L)(\overline{\nu}_L\gamma^\mu\nu'_L)\\&+ c^{RR}(\overline{u}_R\gamma_\mu c_R)(\overline{\nu}_R\gamma^\mu\nu'_R) 
+c^{LR}(\overline{u}_L\gamma_\mu c_L)(\overline{\nu}_R\gamma^\mu\nu'_R)\\& +c^{RL}(\overline{u}_R\gamma_\mu c_R)(\overline{\nu}_L\gamma^\mu\nu'_L) \
+g^{LL}(\overline{u}_Lc_R)(\overline{\nu}_L\nu'_R) \\& +g^{RR}(\overline{u}_Rc_L)(\overline{\nu}_R\nu'_L) 
+g^{LR}(\overline{u}_Lc_R)(\overline{\nu}_R\nu'_L) \\&+ g^{RL}(\overline{u}_Rc_L)(\overline{\nu}_L\nu'_R) 
+h^{LL}(\overline{u}_L\sigma^{\mu\nu}c_R)(\overline{\nu}_L\sigma_{\mu\nu}\nu'_R) \\&+ h^{RR}(\overline{u}_R\sigma^{\mu\nu}c_L)(\overline{\nu}_R\sigma_{\mu\nu}\nu'_L)\bigg]+\text{h. c.}.
\end{aligned}
\end{align} 
In Ref. \cite{Bause:2020xzj} right-handed massless neutrinos are considered. 
Also, in Ref. \cite{Faisel:2020php} authors considered  charm meson decays to invisible fermions, which have negligible masses.  In the following, we consider massive right-handed fermions and use further the notation $\nu_R \equiv \chi_R$. 
Following \cite{Dorsner:2016wpm}, we write in Table  \ref{tab:LQ_scenarios} interactions of the coloured scalar  $\bar S_1$ and  $\tilde R_2$ with the  up quarks   and $\tilde R_2$ and  $S_1$ with down quarks.

\begin{table}[h]
\centering
\begin{tabular}{|c|c|}
\hline
Cloured Scalar& Invisible fermion \\
\hline
$S_1 = (\bar 3,1,1/3)$ & $ \bar d_R ^{C\, i} \chi^j S_1$ \\
$\bar S_1= (\bar 3, 1, -2/3)$ & $ \bar u^{C\,i}_R \chi^j \bar S_1$\\
$\tilde R_2= (\bar 3, 2, 1/6)$ & $\bar u_L^i \chi^j  \tilde R_2^{2/3}$ \\
$\tilde R_2= (\bar 3, 2, 1/6)$ & $\bar{d}_{L}^{i}\chi^{j}\tilde{R}_{2}^{-1/3}$\\
\hline 
\end{tabular}
\caption{\label{tab:LQ_scenarios} The coloured scalars $\bar S_1$, $S_1$ and $\tilde R_2$ interactions with invisible fermions and quarks. Here we use only right-handed couplings of $S_1$. Indices $i,j$ refer to quark generations.}
\end{table}
 We concentrate only on coloured scalar and scalar leptoquark due to  difficulties with vector leptoquarks. Namely,  the simplest way to consider vector leptoquarks in an ultra-violet complete theory is when they play the role of gauge bosons. For example, $U_1$ is one of the gauge bosons in some of Pati-Salam unification schemes  \cite{Bordone:2018nbg,DiLuzio:2017vat}. However,  other particles with masses close to  $U_1$ with many new parameters in such theories, making it rather difficult  to use without additional assumptions. 
 
Coloured scalars contributing to transition $c \to u \chi  \bar \chi $ have following Lagrangians, as already anticipated in 
in \cite{Dorsner:2016wpm} 
\begin{equation}
\label{eq:main_b_S_1}
\mathcal L (\bar S_1) \supset \bar{y}^{RR}_{1\,ij}\bar{u}_{R}^{C\,i}\, \chi_{R}^{j} \,\bar{S}_1
+\textrm{h.c.}.
\end{equation} 
\begin{equation}
\mathcal{L} (\tilde R_2) \supset 
(V \tilde{y}^{LR}_2)_{ij} \bar{u}_{L}^{i} \,\chi_{R}^{j} \tilde{R}_{2}^{2/3}+\tilde{y}^{LR}_{2\,ij}\bar{d}_{L}^{i}\,\chi_{R}^{j}\tilde{R}_{2}^{-1/3}+\textrm{h.c.}.
\label{eq:main_t_R_2_a}
\end{equation}
Here, we give only  terms containing interactions of quarks with right-handed $\chi$.  The $S_1$ scalar leptoquark, in principle, might mediate $c\to u \chi \bar \chi$ on the loop level, with one $W$ boson changing  down-like quarks to $u$ and $c$. Obviously, such a loop process is also suppressed by  loop factor $1/(16 \pi^2)$ and $G_F$ making it negligible. 
Also,  due to the right-handed nature of $\chi$, one can immediately see that in the case of $\bar S_1$,  the effective Lagrangian has only one  contribution 
\begin{equation}
{\mathcal L}_{\text{eff} } = {\sqrt 2} G_F c^{RR} \left( \bar u_R \gamma_\mu c_R \right) \left( \bar \chi_R \gamma^\mu \chi_R \right),
\label{effS}
\end{equation}
with
\begin{equation}
c^{RR} = \frac{v^2}{2 M_{\bar S_1}^2} \bar{y}^{RR}_{1\,c \chi } \, \bar{y}^{RR\ast}_{1\,u \chi }.
\label{cRR}
\end{equation}
In the case of $\tilde R_2$ 
\begin{equation}
{\mathcal L}_{\text{eff} }= {\sqrt 2} G_F c^{LR} \left( \bar u_L \gamma_\mu c_L \right) \left( \bar \chi_R \gamma \chi_R \right),
\label{effR}
\end{equation}
with
\begin{equation}
c^{LR} = - \frac{v^2}{2 M_{\tilde{R}_2}^2} \left( V\tilde{y}_2^{LR}\right)_{u \chi } \left( V\tilde{y}_2^{LR}\right)_{c \chi }^\ast.
\label{cLR}
\end{equation}
For the mass of $\chi$, kinematically allowed, in the $c\to u \chi \bar \chi$ decay, one can relate this amplitude to $b\to s \chi \bar \chi$ or in $s \to d \chi \bar \chi$. However, it was found \cite{Ruggiero} that the  experimental  rates  for $K \to \pi \nu \bar \nu$ are very close to the SM rate \cite{Buras:2015qea}, leaving  very little room for  NP contributions. Therefore, we avoid this kinematic region and consider mass of $\chi$ to be $m_\chi \geq (m_K -m_\pi)/2$, while the charm decays allow  $m_\chi \leq (m_D - m_\pi)/2$. For our further study it is very important that $\chi$ is a  weak singlet and therefore  LHC searches of high-$p_T$ lepton tails \cite{Angelescu:2020uug,Fuentes-Martin:2020lea} are not applicable for the constraints of interactions in the cases we consider. 
However, further study of final states containing mono-jets and missing at LHC and future High luminosity colliders will shed more light on these processes.

\section{$\bar S_1$ in $c \to u \chi  \bar \chi $}

Due to its quantum numbers, the coloured scalar $\bar S_1$  and $\chi$ can interact only with up-like quarks.
Most generally, the number of $\chi$'s  can be three and the matrix $y_{1\,ij}^{RR}$ can have $9 \times 2$ parameters. Here, we consider one $\chi$, that can couple to both $u$ and $c$ 
quarks. 
These two couplings might enter in amplitudes for processes with down-like quarks at loop-level, as discussed in \cite{Fajfer:2020tqf}.
Obviously,  due to the right-handed nature of $\chi$, one can immediately see that in the case of $\bar S_1$,  
the effective Lagrangian has only the contribution
\begin{equation}
\mathcal{L}_{\text{eff}} =\sqrt{2}G_F  \frac{v^2}{2 M^2}  \bar y_{1\,c \chi}^{RR} \bar y^{RR*}_{1\, u \chi} (\overline{u}_R\gamma_\mu c_R)(\overline{\chi}_R\gamma^\mu\chi_R). 
\label{RR-Lag}
\end{equation}
First, we discuss constraints from $D^0 -\bar D^0$ mixing and then consider exclusive decays $ D^0 \to \chi \bar \chi$, $ D^0 \to \bar \chi \chi \gamma$, and $D\to \pi \chi \bar \chi$.
The authors of  Ref. \cite{Faisel:2020php} 
considered scalar leptoquarks allowing each up-quark can couple to different flavour of lepton or right-handed neutrino. In such a way, they avoid constraints from the  $D^0 -\bar D^0$ mixing. 

\subsubsection{Constraints from $D^0 -\bar D^0$} 

The strongest constraints on $\chi$ interactions with $u$ and $c$ comes from the $D^0 -\bar D^0$ oscillations. The interactions in Eqs. (\ref{eq:main_b_S_1}) and (\ref{eq:main_t_R_2_a}) can generate transition $D^0 -\bar D^0$. Coloured scalar $\bar S_1$ contributes to the operator entering  the effective Lagrangian \cite{Fajfer:2015mia,Dorsner:2016wpm} 
\begin{equation}
{\mathcal L}_{\text{eff}}^{Dmix} = -C_6 \left( \bar c\gamma_\mu P_R u\right) \left( \bar c\gamma^\mu P_R u\right),
\label{Dmix}
\end{equation}
with the Wilson coefficient given by
\begin{equation}
C_6 = \frac{1}{64 \pi^2 M_{\bar S_1}^2} \left( \bar{y}^{RR}_{1\,c\chi } \right)^2  \left( \bar{y}^{\overline{RR}\ast}_{1\,u\chi } \right)^2.
\label{C6}
\end{equation}
The standard way to write the hadronic matrix element is $\left<\bar D^0| (\bar u \gamma_\mu P_R c)  (\bar u \gamma^\mu P_R c)| D^0\right> = \frac{2}{3} m_D^2 f_D^2 B_D$, with the bag parameter $B_D(3 \, {\rm GeV})= 0.757(27)(4)$ calculated in the MS scheme, which was computed by the lattice QCD \cite{Carrasco:2015pra} and the D meson decay constant defined as $\left<0| \bar u \gamma_\mu \gamma_5 c)| D(p)\right>= i f_D p_\mu$, with  $f_D = 0.2042$ GeV \cite{Zyla:2020zbs}.
Due to large nonperturbative contributions, the SM contribution is not well known. Therefore, in the
absence of CP violation, the robust bound on the product of the couplings can be obtained by requiring that the
mixing frequency should be smaller than the world average $x=2|M_{12}|/\Gamma = (0.43^{+0.10}_{-0.11})\%$ by HFLAV \cite{Amhis:2019ckw}. 
The bound on this Wilson coefficient can be derived following \cite{Fajfer:2020tqf,Fajfer:2008tm}
\begin{equation}
\left| r C_6 (M_{\bar S_1} )\right| \frac{2 m_D f_D^2 B_D}{3 \Gamma_D} < x,
\label{C6-con}
\end{equation}
with  a renormalisation factor $r=0.76$ due to running of $C_6$ from scale $M_{\bar S_1} \simeq 1.5$ TeV down to $3$ GeV. One can derive $|C_6 |< 2.3\times 10^{-13}$ ${\rm GeV}^{-2}$ or
\begin{equation} 
\left|  \bar{y}^{RR}_{1\,c\chi } \, \bar{y}^{RR\ast}_{1\,u\chi } \right| < 1.2 \times 10^{-5} M_{\bar S_1}/ GeV.
\label{yyD}
\end{equation}
\begin{equation}
c^{RR} < \frac{0.363 \, {\rm GeV}}{M_{\bar S_1} ({\rm GeV})}.
\label{cRR-D}
\end{equation}

\begin{figure}[!hbp]
\centering
\includegraphics[scale=0.5]{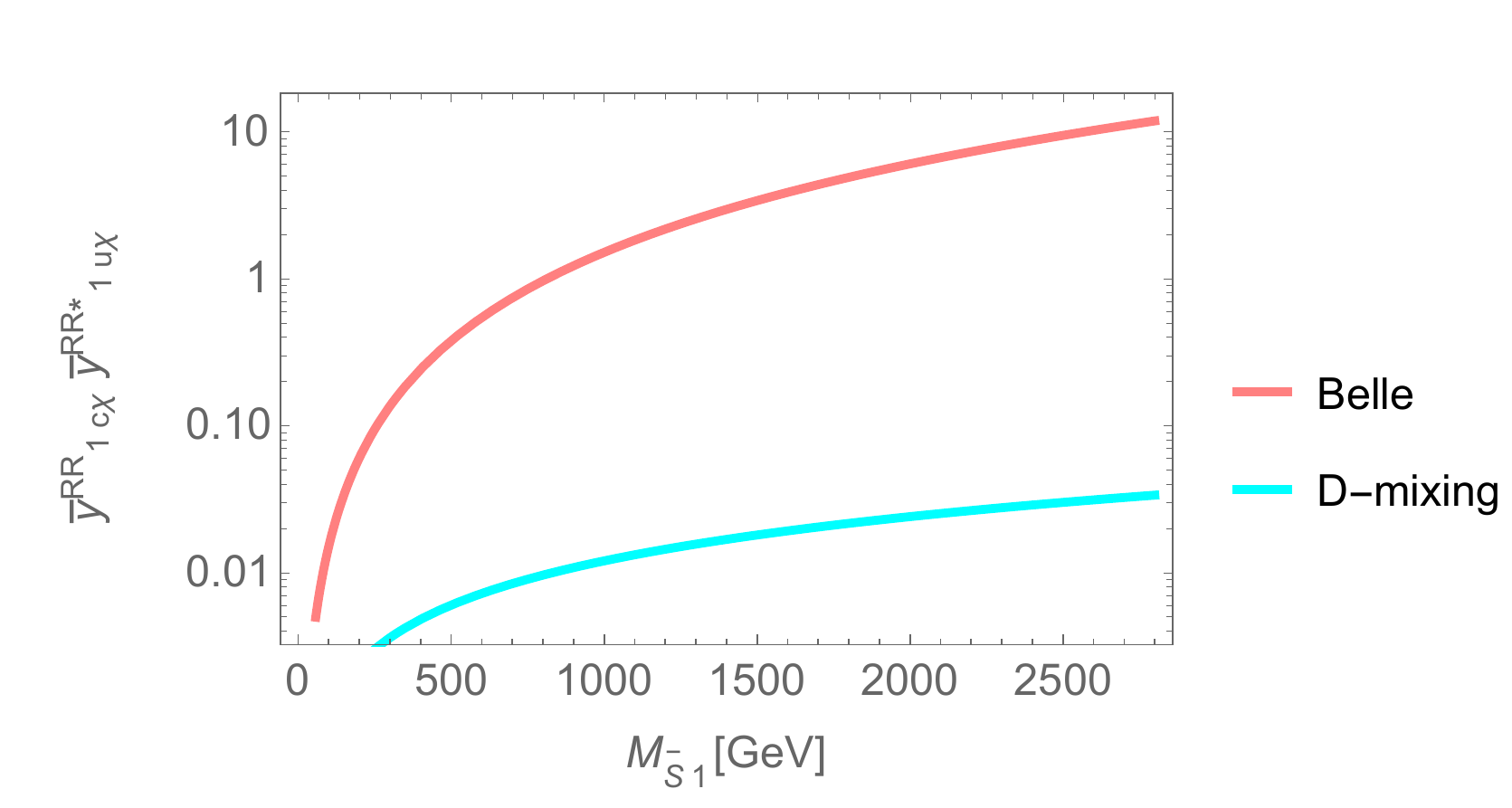}
\caption{ \label{Fig.1} The product of Yukawa couplings $\left|  \bar{y}^{RR}_{1\,c\chi } \, \bar{y}^{RR\ast}_{1\,u\chi } \right| $  as a function of the $\bar S_1$ mass. The pink line
denotes the bound derived from Belle result \cite{Lai:2016uvj}, while the turquoise one is obtained with the bound from $D^0 - \bar D^0$ oscillations.
}
\end{figure}

\subsubsection{$ D^0 \to \chi \bar \chi$}

\begin{figure}[H]
\centering
\includegraphics[scale=0.4]{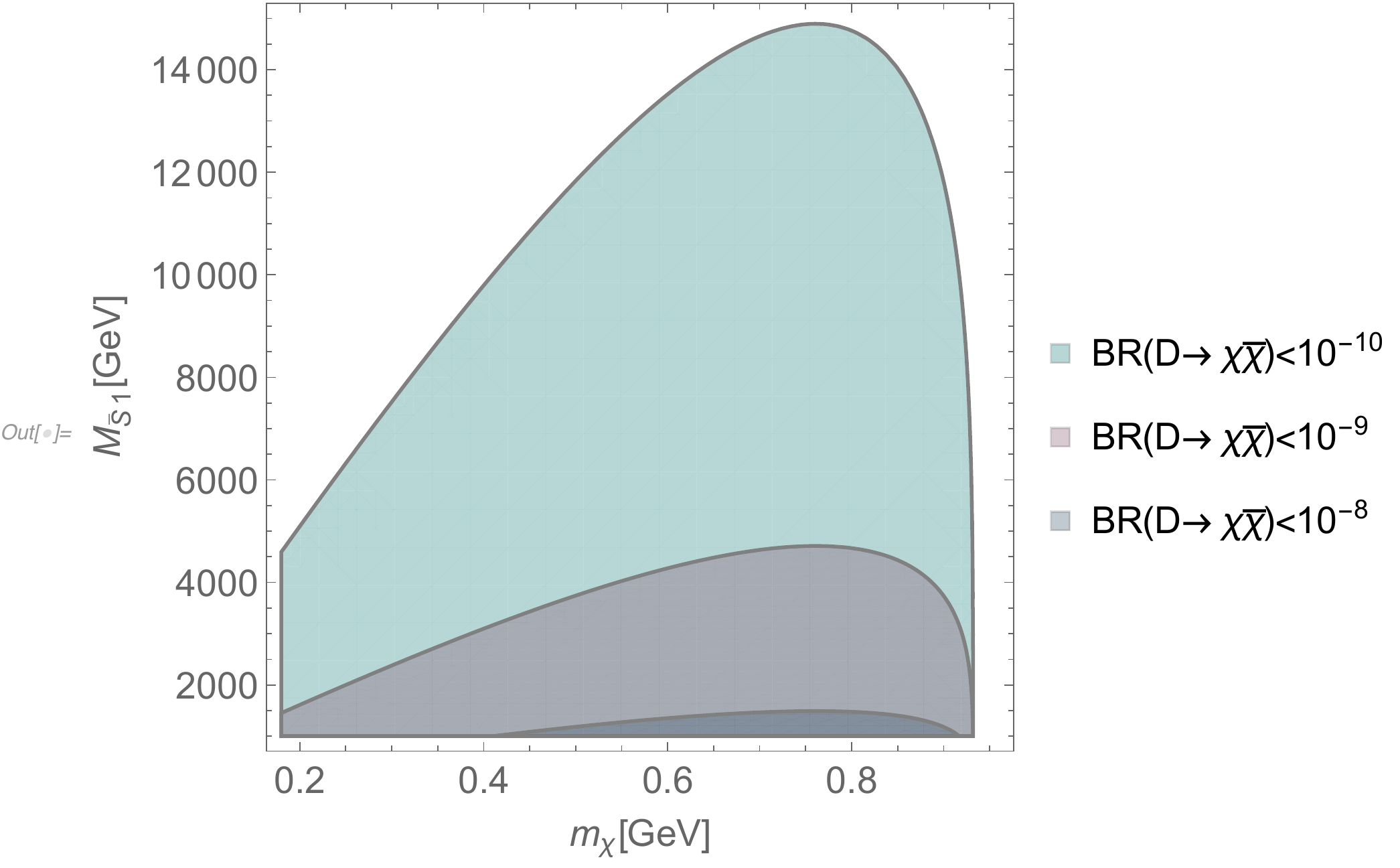}
\caption{ \label{Fig.2} The allowed mass region for $\bar S_1$ in the range  $(m_K -m_\pi)/2 < m_\chi < (m_D -m_\pi)/2 $. The regions are obtained assuming 
${\mathcal B} ( D^0 \to \chi \bar \chi ) < 10^{-10}$, $10^{-9}$ and $10^{-8}$, for the product $\left|  \bar{y}^{RR}_{1\,c\chi } \, \bar{y}^{RR\ast}_{1\,u\chi } \right|  =1$.}
\end{figure}

The amplitude for this process can be  written as 
\begin{equation}
{\mathcal M} (D^0 \to \chi \bar \chi)=  \frac{\sqrt 2}{2} G_F f_D c^{RR} m_\chi \bar u_\chi (p_1) \gamma_5 v_\chi (p_2),
\label{AD2}
\end{equation}
 giving the branching ratio
\begin{equation}
{\mathcal B} ( D^0 \to \chi \bar \chi ) = \frac{1}{\Gamma_D} \frac{G_F^2 f_D^2 m_D}{16 \pi} \left|c^{RR}\right|^2 m_\chi^2 \sqrt{1 - \frac{4 m_\chi^2}{m_D^2}}.
\label{BRD2}
\end{equation}
Using Belle bound ${\mathcal B} ( D^0 \to \chi \bar \chi )  < 9.4 \times 10^{-5}$ \cite{Lai:2016uvj},  one can find easily the  bound on Wilson coefficient
$\left|c^{RR}\right|_{Belle} < 0.046$. 
This value is derived for the mass $m_\chi = 0.8$ GeV. 
We analyse the dependence on the mass of $\bar S_1$, allowing the mass of $\chi$ to be $(m_K -m_\pi)/2 < m_\chi < (m_D -m_\pi)/2 $, and  assume the branching ratio for  
${\mathcal B} ( D^0 \to \chi \bar \chi ) < 10^{-10}$, $10^{-9}$ and $10^{-8}$, with $ \left|  \bar{y}^{RR}_{1\,c\chi } \, \bar{y}^{RR\ast}_{1\,u\chi } \right| =1$. We present our result in Fig. \ref{Fig.2} and find  that mass of $\bar S_1$, using these reasonable assumptions, can be within LHC reach.
\begin{table}[h]
\centering
\begin{tabular}{|c|c|} 
\hline
$m_\chi$ (GeV)& ${\mathcal B} ( D^0 \to \chi \bar \chi)_{D-\bar D}$\\
\hline
0.18 & $<1.1\times 10^{-9} $\\
0.50 & $<7.4\times 10^{-9}$\\
0.80 & $ <1.1\times 10^{-8} $\\
\hline 
\end{tabular}
\caption{\label{BR-Drad} Branching ratios for  ${\mathcal B} ( D^0 \to \chi \bar \chi) $ for three selected values of $m_\chi$. The constraints from the $D^0 - \bar D^0$ mixing is used, with $c^{RR}\leq 3.63 \times 10^{-4}$, assuming  $M_{\bar S_1} = 1000$  GeV.}
\end{table}

\subsubsection{$ D^0 \to \chi \bar \chi \gamma$}

The authors of Ref. \cite{Badin:2010uh} suggested, that  the helicity suppression, present in the $ D^0 \to \chi \bar \chi $ amplitude for $m_\chi =0$,  is  lifted by an additional photon in the final state and therefore $ D^0 \to \chi \bar \chi \gamma$  might bring additional information on detection of invisibles in the final state.
They found that the branching decay is 
\begin{eqnarray}
&&{\mathcal B} ( D^0 \to \chi \bar \chi \gamma) = \frac{G_F ^2F_{DQ} ^2f_D^2  |c^{RR}|^2  m_D^2\alpha}{1152 \pi^2 \Gamma_D \sqrt{1-4 x_\chi^2}}  Y(x_\chi).\nonumber\\
\label{BRDrad}
\end{eqnarray}
In the above equations $x_\chi = m_\chi/ m_D$, $F_{DQ} =2/3 (-1/(m_D-m_c) +1/m_c)$, $f_D= 0.2042$ GeV \cite{Zyla:2020zbs} and $Y(x_\chi)$ is given in Appendix. Coefficient $c^{RR}$ is  constrained by Eq. (\ref{cRR-D}). 
\begin{table}[h]
\centering
\begin{tabular}{|c|c|c|} 
\hline
$m_\chi$ (GeV)& ${\mathcal B} ( D^0 \to \chi \bar \chi \gamma)_{D-\bar D}$& ${\mathcal B} ( D^0 \to \chi \bar \chi \gamma)_{Belle}$\\
\hline
0.18 & $<2.1\times 10^{-11} $&  $<1.3 \times 10^{-7} $\\
0.50 & $<6.9\times 10^{-12} $&  $<6.3\times 10^{-9} $\\
0.80 & $ <8.4\times 10^{-14} $&  $<2.2\times 10^{-10} $\\
\hline 
\end{tabular}
\caption{\label{BR-Drad} Bounds on the branching ratio for  ${\mathcal B} ( D^0 \to \chi \bar \chi \gamma) $. In the second column the constraints from the $D^0 - \bar D^0$ mixing is used, assuming  $M_{\bar S_1} = 1000$  GeV. In the third column Belle bound ${\mathcal B} ( D^0 \to \slashed{E})< 9.4 \times 10^{-5}$ is used.}
\end{table}
Comparing these results with the SM result presented in Ref. \cite {Badin:2010uh} ${\mathcal B} ( D^0 \to  \nu \bar \nu \gamma)_{SM} =3.96 \times 10^{-14}$, we see that the existing Belle bound allows significant branching ratio, while the bounds from the $D^0 - \bar D^0$ mixing, for larger values of  $m_\chi$, lead to the branching ratio to be  close to the SM results. Due to the mass of $\chi$, the photon energy  can be in the range $0\leq E_\gamma\leq (m_D^2 -m_\chi^2)/(2m_D)$,  which in principle would distinguish the SM contribution from the contributions with massive invisible fermions.

\subsubsection{$D\to \pi \chi \bar \chi$}

The rare charm decays due to GIM-mechanism cancellation are usually dominated by long distance contributions.    Long distance contributions to exclusive decay channel $D \to \pi \nu \bar \nu$ were considered in Ref. \cite{Burdman:2001tf}.  For example, the branching ratio $BR(D^+\to \pi^+  \rho^0 \to \pi^+ \nu \bar \nu )< 5 \times 10^{-16}$. The authors of \cite{Burdman:2001tf}  discussed another possibility $D^+ \to \tau^+ \nu \to \pi ^+\bar \nu \nu$ and found that  the branching ratio  should be smaller than $1.8\times 10^{-16}$. 
An interesting study of these effects was done in Ref. \cite{Kamenik:2009kc}, implying that in order to avoid these effects one should make cuts in the invariant $\chi \bar \chi$ mass square, $q_{cut}^2< (m_\tau^2 -m_\pi^2)(m_D^2- m_\tau^2)/m_\tau^2$.  

The amplitude for $D\to \pi \chi \bar \chi$ can be  written as 
\begin{eqnarray}
{\cal M}(D\to \pi \chi \bar \chi)&=& {\sqrt 2} G_F c^{RR} \bar u_\chi (p_1) \gamma_\mu P_R v_\chi(p_2)
\nonumber\\
&\enspace&\left<\pi (k) | \bar u \gamma^\mu P_R| D (p)\right> ,\label{A-Dpi}
\end{eqnarray}
with the standard form-factors definition
\begin{eqnarray} 
&&\left<\pi (k) | \bar u \gamma^\mu (1\pm \gamma_5)| D (p)\right> = f_+ (q^2) \left[ (p + k)^\mu - \frac{m_D^2- m^2_\pi}{q^2} q^\mu \right]\nonumber\\
&&+ f_0 (q^2) \frac{m^2_D-m^2_\pi}{q^2}  q^\mu,
\label{FF}
\end{eqnarray}
with $q=p-k$. We follow the  update of the form-factors in Ref.  \cite{Fleischer:2019wlx}. 
This enables us to write the amplitudes in  the form given in Ref. \cite{Fajfer:2015mia}
\begin{eqnarray}
&& \mathcal{M} (D(p) \to \pi(k) \chi(p_1) \bar \chi (p_2))= \frac{\sqrt{2}}{2}G_F [V(q^2) \bar{u}_\chi (p_1)\slashed{p}v_\chi (p_2)\nonumber\\
&&+ A(q^2)\bar{u}_\chi (p_1)\slashed{p}\gamma_5v_\chi (p_2) + P(q^2)\bar{u}_\chi (p_1)\gamma_5 v_\chi(p_2) ],
\label{Dpi-inv}
\end{eqnarray}
with the following definitions 

\begin{equation}
\begin{aligned}
V(q^2)&=A(q^2)\equiv c^{RR}f_+(q^2)\\
P(q^2)&\equiv -c^{RR}m_\chi \left[f_+(q^2)-\frac{m_D^2-m_\pi^2}{q^2}(f_0(q^2)-f_+(q^2))\right].
\end{aligned}
\end{equation}

We can the differential decay rate distribution as 
\begin{equation}\label{dBRdq}
\frac{d\mathcal{B}(D\rightarrow\pi\bar{\chi}\chi)}{dq^2}=\frac{1}{\Gamma_D} N \lambda^{1/2}\beta\left[ 2a(q^2) + \frac{2}{3}c(q^2)\right].
\end{equation}
with notation $\lambda \equiv \lambda(m_D^2,m_\pi^2,q^2)$,  ($\lambda(x,y,z)=(x+y+z)^2-4(xy+yz+zx)$), $\beta =\sqrt{1-4m_\chi^2/q^2}$ and $N=\frac{G_F^2 }{64(2\pi)^3 m_D^3}$. Note that in case of charged charm meson there is a multiplication by 2 in the differential decay rate compared to neutral $D$.

\begin{figure}[H]
\centering
\includegraphics[scale=0.5]{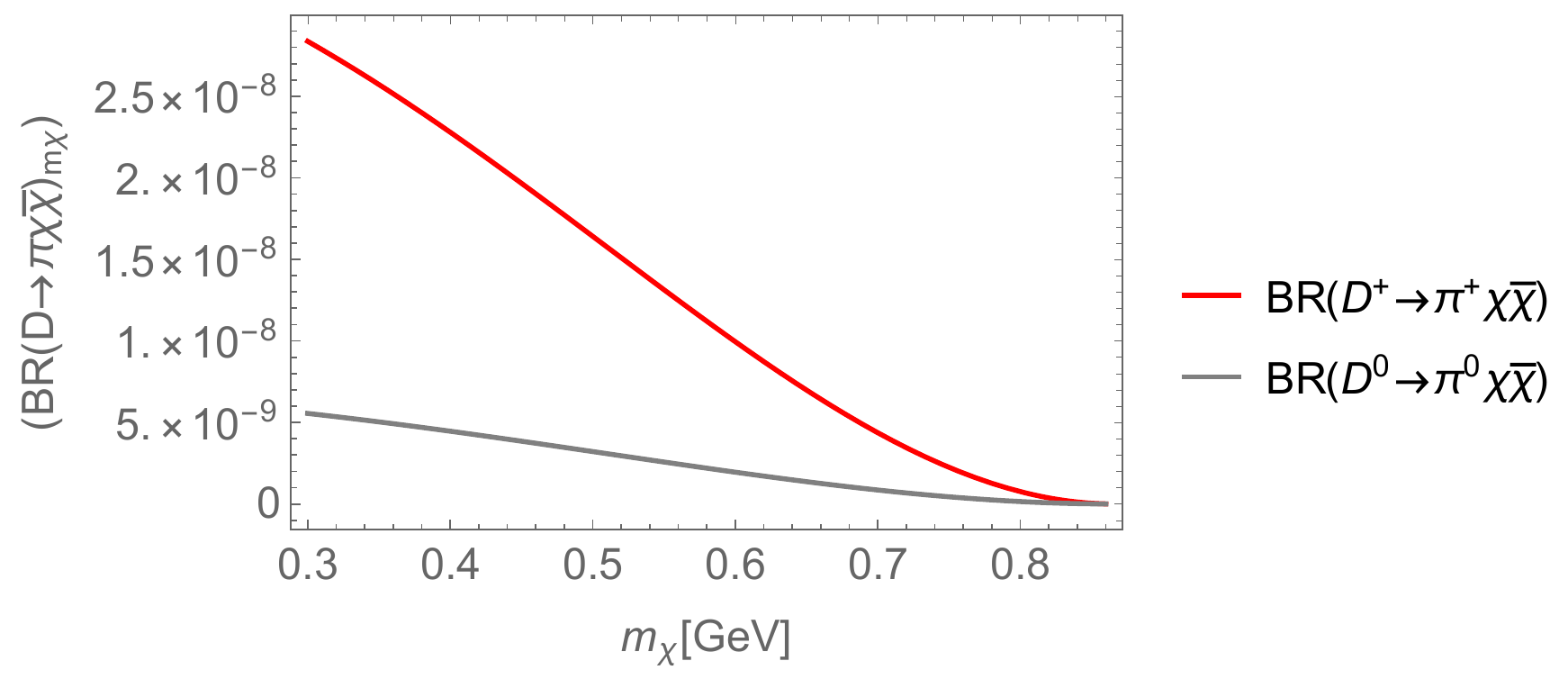} 
\caption{ \label{Fig4} Branching fraction for $D^+\to \pi^+\chi \bar \chi $ and $D^0\to \pi^0\chi \bar \chi $ as a function of $m_\chi$.}
\end{figure}
The integration bounds should be  $4 m_\chi^2 \leq q^2 \leq (m_D-m_\pi)^2$ in the case of $m_\chi =0.5,\, 0.8$, while instead of $m_\chi =0.18$ GeV,  $q^2_{cut}$ is used from Ref. \cite{Kamenik:2009kc}, giving the lowest  mass of the invisibles should be searched in the region $m_\chi \geq \sqrt{q^2_{cut}/ 4}\simeq 0.29$ GeV. This enables us to avoid the region in which the effects of the long distance dynamics dominates.
\begin{table}[h]
\centering
\begin{tabular}{|c|c|c|} 
\hline
$m_\chi$ (GeV)& ${\mathcal B} ( D^0 \to \pi^0 \chi \bar \chi )_{D-\bar D}$& ${\mathcal B} ( D^+\to \pi^+\chi \bar \chi )_{D-\bar D}$\\
\hline
0.18& $<5.9\times 10^{-9} $&  $<3.0 \times 10^{-8} $\\
0.50 & $<3.2\times 10^{-9} $&  $<1.6 \times 10^{-8} $\\
0.80 & $ <1.5 \times 10^{-10} $&  $<7.6\times 10^{-10} $\\
\hline 
\end{tabular}
\caption{\label{BR-Drad} Branching ratios for  ${\mathcal B} ( D \to \pi\chi \bar \chi) $. In the second and the third columns the constraint from the $D^0 - \bar D^0$ mixing is used, assuming the mass of $M_{\bar S_1} = 1000$  GeV.  In the case $m_\chi =0.18$, the cut in integration variable is done by taking $q_{cut}^2$, as described in the text.}
\end{table}
One can use the Belle bound  \cite{Lai:2016uvj} for $\mathcal B(D\to \slashed{E})$ and determine $c^{RR}$ from $D^0\to \chi \bar \chi$ for each $\chi$  mass. 
We obtain $\mathcal B ( D^0 \to \pi^0 \chi \bar \chi )_{Belle} \le 4.9 \times 10^{-4},\, 4.0 \times 10^{-5}, \, 1.2 \times 10^{-6}$  and  $\mathcal B ( D^+ \to \pi^+ \chi \bar \chi )_{Belle} \le 2.5 \times 10^{-3},\, 2.1 \times 10^{-4}, \,  6.1  \times 10^{-6}$ for $m_\chi =0.18,\, 0.5, \, 0.8$ GeV respectively. 
Obviously, the current Belle bound used in the Wilson coefficient leads to the significant increase of the branching ratios for both decay modes. Although the charm meson mixing is very constraining for the relevant couplings, the calculated branching ratios reaching the order $10^{-8}$ might be possible to observe at the future tau-charm factories and Belle II experiment.

\section{$\tilde R_2$ in $c\to u \chi \bar \chi$}

The $\tilde R_2$ leptoquark  is a weak doublet and it interacts with quark doublets (\ref{eq:main_t_R_2_a}). Therefore, the appropriate couplings, $\tilde y^{LR}_{2\,s \chi}\, \tilde y_{2\, b \chi}^{LR\ast}$ can be constrained from the  $b\to s \chi \bar \chi$ and 
$s\to d \chi \bar \chi$ decays, as well as from  observables coming from the  $B_s - \bar B_s$, $B_d - \bar B_d$, $K^0 -\bar K^0$ oscillations  as in \cite{Fajfer:2020tqf}. 
We consider the most constraining  bounds coming from decays $B\to K \slashed{E}$ and from the oscillations of $B_s - \bar B_s$,  relevant for the  $\chi$ mass region  $(m_K-m_\pi)/2 < m_\chi < (m_D - m_\pi)/2$.   The decay $B\to K \slashed{E}$ was recently studied by the authors of Ref. \cite{Li:2020dpc}. They pointed out  that current bound on  the rate $B\to K \slashed{E}$ when the SM branching ratio for $B\to K \nu \bar \nu$ is subtracted from the experimental bound on $\mathcal B( B^+ \to K^+  \slashed{E})$ is the most constraining.  They derived 
${\mathcal B} (B\to K\slashed{E})< 9.7 \times 10^{-6}$as the strongest bound among $B \to H_s  \slashed{E}$ ($H_s$ is a hadron containing the  $s$ quark). 

\subsubsection{Constraints from $B\to K \slashed{E}$ and $B_s - \bar B_s$ oscillations}

The amplitude for $B\to K \chi \bar \chi$  can be written as 
\begin{eqnarray}
{\cal M}(B\to K \chi \bar \chi)&=& {\sqrt 2} G_F c^{LR}_{B} \bar u_\chi (p_1) \gamma_\mu P_R v_\chi(p_2)
\nonumber\\
&\enspace&\left<K (k) | \bar u \gamma^\mu P_L| B (p)\right> .\label{A-Dpi}
\end{eqnarray}
In the case of Wilson coefficient $c^{LR}_B$ it is easy to find  \cite{Dorsner:2016wpm}
\begin{equation}
c^{LR}_{B} =- \frac{v^2}{2 M^2_{\tilde R_2}} \tilde  y^{LR}_{2\,s \chi}  \tilde y^{LR\ast}_{2\,b\chi} .
\label{PCLR}
\end{equation}
The integration over the phase space depends on the mass of $m_\chi$ we chose. Here we can choose a mass $\chi$, which we used in D decays $(m_K -m_\pi)/2 < m_\chi < (m_D -m_\pi)/2 $. 
The bounds on the Wilson coefficient  in Eq. (\ref{PCLR}) are following $ |c^{LR}_{B}|<3.3 \times10^{-4}$, $ <4.9\times10^{-4}$ and $<9.1\times10^{-4}$ for $m_\chi= 0.18,\, 0.50,\, 0.80$ GeV.

There are two box diagrams with $\chi$  within the box contributing to the  $B_s - \bar B_s$ oscillations. The contribution of $\tilde R_2$ box diagrams to the effective Lagrangian for the $B_s - \bar B_s$ oscillation is
 \begin{eqnarray}
&& {\cal L}^{NP}_{\Delta B=2} =-\frac{1 }{128 \, \pi^2} \frac{ \left( \tilde y^{LR}_{2\,s \chi }\right)^2 \left(  \tilde y^{LR\,\ast  }_{2\, b\chi }\right)^{2} }{M_{\tilde R_2}^2} \nonumber\\
&& \times \left(\bar s \gamma_\mu P_R b \right) \, \left(\bar s \gamma^\mu P_R b \right).
 \label{deltamBs-S1}
 \end{eqnarray}

\begin{figure}[!hbp]
\centering
\includegraphics[scale=0.48]{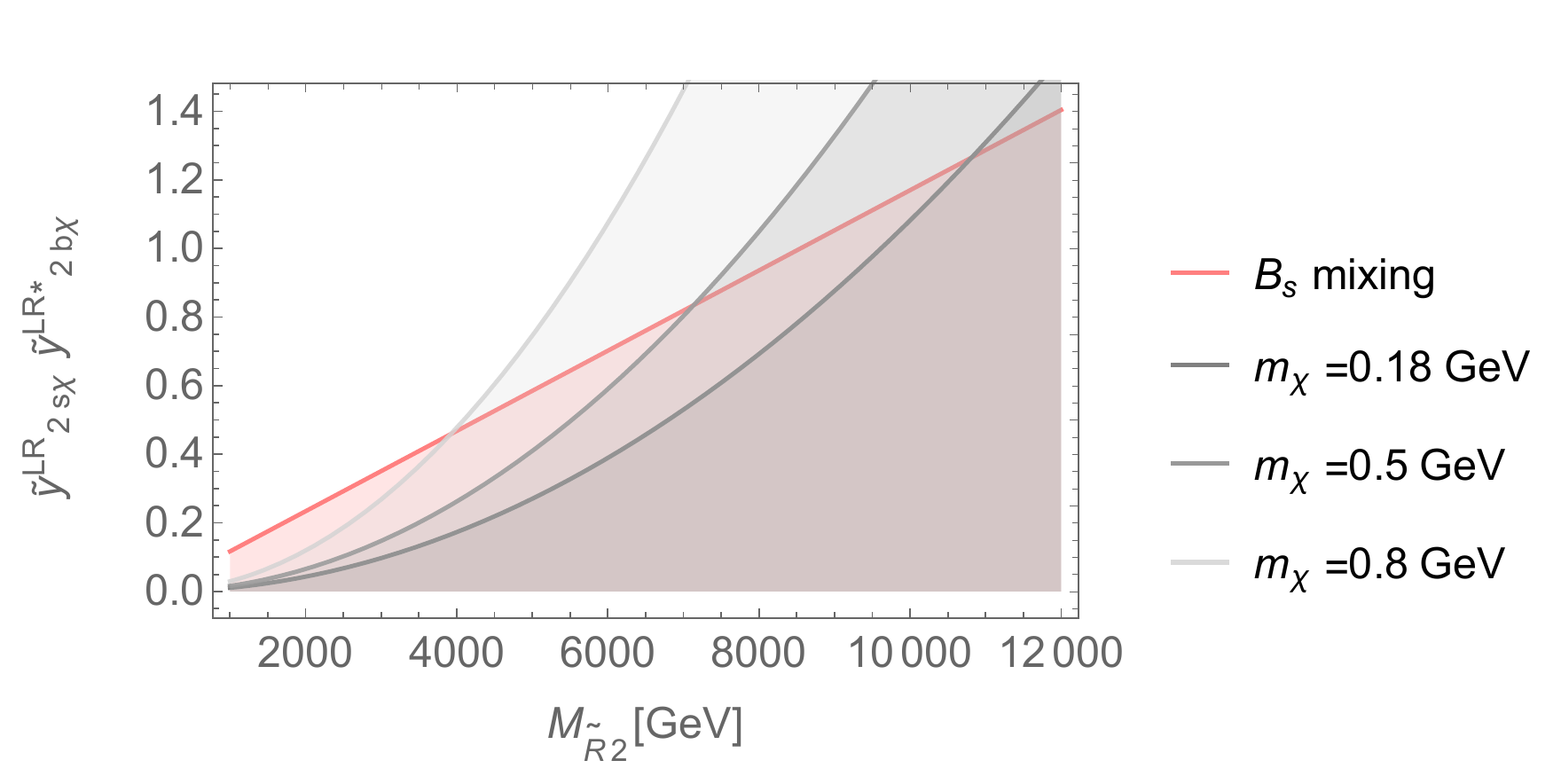}
\caption{ \label{Fig.4} The allowed mass for $\tilde R_2$. Constraints are derived from the $B_s^0 -\bar B_s^0$ mixing and from the bound 
$\mathcal B(B^+ \to K^+ \slashed{E})< 9.7\times 10^{-6}$ for $m_\chi =0.18,\, 0.5,\, 0.8$ GeV. }
\end{figure}

We can understand this result in terms of the recent study of new physics in the $B_s - \bar B_s$ oscillation in \cite{DiLuzio:2019jyq}. 
 The authors of \cite{DiLuzio:2019jyq} introduced the following notation of the New Physics (NP) contribution containing the right-handed operators as
 \begin{equation} 
 {\cal L}_{\Delta B=2}^{NP} \supset -\frac{4 G_F}{\sqrt 2} (V_{tb}V_{ts}^*)^2 C_{bs}^{RR} \left(\bar s \gamma_\mu P_R b \right) \, \left( \bar s \gamma^\mu P_R b \right).
 \label{Bs-NP}
 \end{equation}
 Following their notation, one can write the modification of the SM contribution by the NP as in Ref. \cite{DiLuzio:2019jyq}
 \begin{equation}
 \frac{\Delta M_s^{SM+NP}}{\Delta M_s^{SM}}= \left|1 + \frac{\eta^{6/23}}{R_{loop}^{SM}} \, C_{bs}^{RR} \right|
 \label{rationSM-NP}
 \end{equation}
 They found that $R_{loop}^{SM} = (1.31 \pm0.010)\times 10^{-3}$ and $\eta = \alpha_s(\mu_{NP}) /\alpha_s(\mu_b)$. Relying on the Lattice QCD results of the two collaborations FNAL/MILC \cite{Bazavov:2016nty}, HPQCD \cite{Dowdall:2019bea}, the FLAG averaging group \cite{Aoki:2019cca} published following results, which we use in our calculations
 \begin{eqnarray}
\Delta M_s^{FLAG 2019} &= &(20.1^{+1.2}_{-1.6}) \, ps^{-1} = (1.13^{+0.07}_{-0.09}) \, \Delta M_s^{exp} ,\nonumber\\
 \label{MBa,e}
 \end{eqnarray}
From these results, one can easily determine bound 
 \begin{equation}
 \Bigg|\frac{ \left( \tilde y^{LR}_{2\, s\chi}\right)^2 \left( \tilde y^{LR\,\ast }_{2\, b \chi }\right)^{2} }{M_{\tilde R_2}^2}\Bigg| \leq 1.39 \times 10^{-8}\, {\rm GeV}^{-2},
\label{Bscoef}
\end{equation}
The same couplings $\tilde y^{LR}_{2\, s\chi} \, \tilde y^{LR\, \ast }_{2\, b \chi }$ enter in the $D^0-\bar D^0$ mixing (\ref{Dmix}) and condition  (\ref{yyD}), and one can derive
\begin{eqnarray}
&&\left[\left( V_{us} \tilde y^{LR}_{2\, s\chi} \right) \,\left(V_{cb} \tilde y^{LR}_{2\, b \chi }\right)^\ast   +
\left(V_{cs} \tilde y^{LR}_{2\, s\chi} \right) \,\left( V_{ub} \tilde y^{LR}_{2\, b \chi } \right)^\ast  \right]\Bigg|_{D-\bar D}\nonumber\\
&& < 1.2 \times 10^{-5} M_{\tilde R_2}/{\rm GeV}.
\label{D-mix-R2}
\end{eqnarray}
The bound on coefficients in (\ref{Bscoef}) lead to  the one order of magnitude stronger constraint then one in (\ref{D-mix-R2}), $ \tilde y^{LR}_{2\, s\chi} \, \tilde y^{LR\,\ast}_{2\, b \chi }<1.58 \times 10^{-6} M_{\tilde R_2}/{\rm GeV}$. In our numerical calculations we use this bound and  do not specify the mass of $\tilde R_2$. However, 
one can combine these constraints and determined the $\tilde R_2$ mass, which can satisfy both conditions. 
In Fig. (\ref{Fig.4})  we present dependence of the couplings $\tilde y^{LR}_{2\, s\chi}\tilde y^{LR\,\ast}_{2\, b \chi }$ as a function of mass $M_{\tilde R_2}$ for masses  using constrain from $B_s^0 - \bar B_s^0$ mixing and from the bound $\mathcal B(B^+ \to K^+ \slashed{E})< 9.7\times 10^{-6}$ for $m_\chi =0.18,\, 0.5,\, 0.8$ GeV. 
From Fig. \ref{Fig.4} we see that the largest mass of $\tilde R_2$, which satisfies both conditions  is $M_{\tilde R_2} \simeq 4400,\, 7100,\, 10800$ GeV for the masses $m_\chi = 0.8,\, 0.5,\, 0.18$ GeV respectively.  All  $\tilde R_2$   masses below these limiting values are allowed and interestingly, they are within LHC reach.

\subsubsection{  $\tilde R_2$ in ${\mathcal B} ( D^0 \to  \chi \bar \chi)$, ${\mathcal B} ( D^0 \to  \chi \bar \chi \gamma)$ and ${\mathcal B} (D^+ \to \pi^+ \chi \bar \chi)$}
Using the same expressions as in the previous section, we calculate branching ratios for 
$D^0 \to  \chi \bar \chi$, $D^0 \to  \chi \bar \chi \gamma$  and present them in Table  \ref{BR-dD}. The results for  $D \to  \pi\chi \bar \chi$ are presented in Table \ref{BRR-Dpi}. The Wilson coefficient $c^{LR}_D$ is obtained using the constraint from $B\to K{\it missing\, energy}$. For $m_\chi=0.18, \, 0.5, \, 0.8$ GeV  they are $c^{LR}_{D} =|(V_{us} V_{cb}^\ast + V_{cs} V_{ub}^\ast) c^{LR}_B|=4.4\times 10^{-6}, \,6.6 \times 10^{-6},\, 1.2 \times 10^{-5}$. 
\begin{table}[h]
\centering
\begin{tabular}{|c|c|c|}
\hline
$m_\chi$ (GeV)& ${\mathcal B} ( D^0 \to  \chi \bar \chi)$ &${\mathcal B} ( D^0 \to  \chi \bar \chi \gamma)$\\
\hline
$0.18$ & $<1.6
\times 10^{-13}$&  $<1.9\times 10^{-15}$ \\
$0.50$ & $<2.4\times 10^{-12}$&  $<1.4\times 10^{-15}$ \\
$0.80 $& $ <1.3\times 10^{-11}$&  $<2.7\times 10^{-16}$\\
\hline 
\end{tabular}
\caption{\label{BR-dD} Branching ratios for ${\mathcal B} ( D^0 \to  \chi \bar \chi)$ and ${\mathcal B} ( D^0 \to  \chi \bar \chi \gamma)$. 
The bounds on the Wilson coefficient $c^{LR}_D$ derived from the ${\mathcal B} (B\to K\slashed{E})< 9.7 \times 10^{-6}$ for selected masses of $\chi$ from the range  
$(m_K -m_\pi)/2 < m_\chi < (m_D -m_\pi)/2 $.}
\end{table}
\begin{table}[h]
\centering
\begin{tabular}{|c|c|c|}
\hline
$m_\chi$ (GeV)& ${\mathcal B} (D^0\to \pi^0 \chi \bar \chi)$&${\mathcal B} (D^+ \to \pi^+\chi \bar \chi)$\\
\hline
$0.18$ &   $<8.7\times 10^{-13}$ & $<4.5\times 10^{-12}$\\
$0.50$ &  $<1.1\times 10^{-12}$ & $<5.4\times 10^{-12}$\\
$0.80 $&  $<1.7\times 10^{-13}$ & $<8.7\times 10^{-13}$\\
\hline 
\end{tabular}
\caption{\label{BRR-Dpi} Branching ratios for ${\mathcal B} (D^0 \to \pi^0 \chi \bar \chi)$ and ${\mathcal B} (D^+ \to \pi^+\chi \bar \chi)$. The bounds on the Wilson coefficient $c^{LR}_D$ derived from  ${\mathcal B} (B\to K\slashed{E})< 9.7 \times 10^{-6}$. In the case $m_\chi =0.18$, the cut in the integration variable is done by taking $q_{cut}^2$, as described in the text. }
\end{table}
Compared with the coloured scalar $\bar S_1$ mediation, the branching ratios for all three decay modes are suppressed for several orders of magnitude, indicating the important role of constraints from $B$ mesons.   
Such suppressed branching ratios of  the all rare charm decays mediated  by $\tilde R_2$  is almost impossible to observe.  On the other hand,  decays of hadrons containing $b$ quarks , mediated  by $\tilde R_2$ a much more suitable for searches of invisible fermions. 

\section{Summary and outlook}

We have presented a study on rare charm decays with invisible massive fermions $\chi$ in the final state.  
The mass of $\chi$ is taken to be in the range $(m_K-m_\pi)/2 <m_\chi < (m_D-m_\pi)/2 $, since the current experimental results on $\mathcal B(K\to \pi \nu \bar \nu)$ are very 
close to the SM result, almost excluding the presence of New Physics. 
We considered two cases with coloured scalar mediators of the up-quarks interaction with $\chi$. The simplest model is one with $\bar S_1= (\bar 3,1, -2/3)$, which couples only to weak up-quark singlets, and the second mediator is $\tilde R_2 =(\bar 3, 2, 1/6)$ which couples to weak quark doublets. 

In the case of $\bar S_1$, the relevant constraint comes from the $D^0 - \bar D^0$ oscillations. 
We have calculated branching ratios for $D^0 \to \chi \bar \chi$, $D^0 \to \chi \bar \chi \gamma $ and $D \to \pi \chi \bar \chi$. The charm meson mixing severely constrain 
the branching ratio $D^0 \to \chi \bar \chi$ in comparison with the experimental result for the branching ratio of $D^0 \to \slashed{E}$.
For our choice of $m_\chi$ the branching ratio for $D^0 \to \chi \bar \chi \gamma $ can be calculated using experimental bound on the rate for $D^0\to \slashed{E}$. In this case, there is an enhancement factor  up to three orders of magnitude smaller, depending on the mass of $\chi$ in comparison with the constraints from the   $D^0 - \bar D^0$ oscillations.
The branching ratios for $D \to \pi \chi \bar \chi$, based on charm mixing constraint, are of the order $10^{-9}- 10^{-7}$, suitable for searches at  future tau-charm factories,  BESIII and Belle II experiments.  

In the case of $\tilde R_2$, for the mass range of $\chi$ relevant for  charm meson rare decays,  we rely on constraints coming from $\mathcal B(B\to K\slashed{E})$ and from the $B_s^0 - \bar B_s^0$ mixing. 
We find that the all three decay modes $ D^0 \to  \chi \bar \chi$, $D^0 \to  \chi \bar \chi \gamma$ and $D^+ \to \pi^+ \chi \bar \chi$ are now having branching ratios for a factor $3-4$ orders of magnitude smaller then in the case of coloured scalar $\bar S_1$ mediation, making them very difficult for the observation. 

Interestingly, the mass of both mediators $\bar S_1$ and $\tilde R_2$ are in the range of LHC reach, and hopefully, searches for mono-jets and missing energy might put constraints on their masses.

\section{Acknowledgment} The work of SF was in part financially supported by the Slovenian Research Agency (research core funding No. P1-0035). The work  of AN was partially supported by the Advanced Grant of European Research Council (ERC) 884719 — FAIME.

\section{Appendix}

\subsection{Phase space factors}
In eq. (\ref{BRDrad}) phase space function $Y(x_\chi)$ is used
\begin{eqnarray}
Y(x_\chi)&=& 1- 2 x_\chi^2 + 3 x_\chi^2 ( 3-6 x_\chi^2 + 4 x_\chi^4) {\sqrt 1-4 x_\chi^2} \nonumber\\
&\times& \log \left( \frac{ 2 x_\chi}{ 1 + \sqrt{1-4 x_\chi^2}}\right) - 11x_\chi^4 + 12 x_\chi^6.
\label{polY}
\end{eqnarray}

In Eq. (\ref{dBRdq}) $a(q^2)$ and $c(q^2)$ are introduced denoting 
\begin{equation}
\begin{aligned} 
a(q^2)&=\frac{\lambda}{2}(|V(q^2)|^2+|A(q^2)|^2)+8 m_\chi^2 m_D^2|A(q^2)|^2\\
+2q^2&|P(q^2)|^2 +4 m _\chi (m_D^2-m_\pi^2+q^2)\text{Re}[A(q^2) P(q^2)^*],\\
c(q^2)&=-\frac{\lambda\beta^2}{2}(|V(q^2)|^2+|A(q^2)|^2).
\end{aligned}
\end{equation}

\subsection{ $ D\to \pi $ form factors }

Following  \cite{Lubicz:2017syv} one can use $z-$expansion with 
\begin{equation}
z= \frac{ \sqrt{ t_+ - q^2} - \sqrt{t_+ - t_0} }{\sqrt{t_+ -q^2} + \sqrt{t_+ - t_0} },
\end{equation} 
with $t_+ = (m_D+m_\pi)^2$ and $t_0 = (m_D+ m_\pi) ( \sqrt m_D - \sqrt m_\pi)$. 
The form factors can be written as
\begin{align}
 f^{D \to \pi}_+(q^2) = \frac{f^{D \to \pi}(0)+c^{D \to \pi}_+ (z-z_0) (1+ \frac{1}{2}(z+z_0))}{1- P_V q^2},\\
 f^{D \to \pi}_0(q^2) = \frac{f^{D \to \pi}(0)+c^{D \to \pi}_0 (z-z_0) (1+ \frac{1}{2}(z+z_0))}{1- P_S q^2},
 \end{align}
where $z_0 = z(0,t_0^{\pi})$. 
The fit parameters are given in Table (\ref{fitparametersZseries}). For the most recent discussion on form-factors see also \cite{Becirevic:2020rzi}.

\begin{table}[h!]
\renewcommand{\arraystretch}{1.2}
\begin{center}
\begin{tabular}{  c c c c c } 
 \hline
$f(0)$ & $c_+$ & $P_V$ (GeV)$^{-2}$ &$c_0$ & $P_S$ (GeV)$^{-2}$\\ 
 \hline
\hline
 0.6117 (354) & -1.985 (347) & 0.1314 (127) &-1.188 (256) & 0.0342 (122)\\ 
 \hline
\end{tabular}
\end{center}
\caption{Fit parameters for $f_0$, $f_+$ in the $z$-series expansion for $D\to \pi$ \cite{Lubicz:2017syv}.}\label{fitparametersZseries}
\end{table}

\subsection{$B\to K$ form factors }

Most recent results are presented in FLAG report \cite{Aoki:2019cca}
\begin{equation}
f_+^{BK} (q^2) = \frac{r_1}{ 1- \frac{q^2}{m_R^2} } + \frac{r_2}{ 1- \frac{q^2}{m_R^2} } , 
\end{equation}

\begin{equation}
f_0^{BK} (q^2) = \frac{r_1}{ 1- \frac{q^2}{m_R^{\prime 2}} }. 
\end{equation}
The parameters are $r_1 =0.162$, $r_2 =0.173$, $m_R=5.41$ GeV and $m_{R^{\prime}}= 6.12$ GeV, as in \cite{Aoki:2019cca}. 
\bibliographystyle{elsarticle-num}
\bibliography{current1-c}

\begin{thebibliography}{10}
\expandafter\ifx\csname url\endcsname\relax
  \def\url#1{\texttt{#1}}\fi
\expandafter\ifx\csname urlprefix\endcsname\relax\def\urlprefix{URL }\fi
\expandafter\ifx\csname href\endcsname\relax
  \def\href#1#2{#2} \def\path#1{#1}\fi

\bibitem{Bause:2020xzj}
R.~Bause, H.~Gisbert, M.~Golz, G.~Hiller, {Rare charm $\boldsymbol{c\to
  u\,\nu\bar{\nu}}$ dineutrino null tests for $\boldsymbol{e^+e^-}$-machines}
  (10 2020).
\newblock \href {http://arxiv.org/abs/2010.02225} {\path{arXiv:2010.02225}}.

\bibitem{Badin:2010uh}
A.~Badin, A.~A. Petrov, {Searching for light Dark Matter in heavy meson
  decays}, Phys. Rev. D 82 (2010) 034005.
\newblock \href {http://arxiv.org/abs/1005.1277} {\path{arXiv:1005.1277}},
  \href {https://doi.org/10.1103/PhysRevD.82.034005}
  {\path{doi:10.1103/PhysRevD.82.034005}}.

\bibitem{Bhattacharya:2018msv}
B.~Bhattacharya, C.~M. Grant, A.~A. Petrov, {Invisible widths of heavy mesons},
  Phys. Rev. D 99~(9) (2019) 093010.
\newblock \href {http://arxiv.org/abs/1809.04606} {\path{arXiv:1809.04606}},
  \href {https://doi.org/10.1103/PhysRevD.99.093010}
  {\path{doi:10.1103/PhysRevD.99.093010}}.

\bibitem{Kou:2018nap}
W.~Altmannshofer, et~al., {The Belle II Physics Book}, PTEP 2019~(12) (2019)
  123C01, [Erratum: PTEP 2020, 029201 (2020)].
\newblock \href {http://arxiv.org/abs/1808.10567} {\path{arXiv:1808.10567}},
  \href {https://doi.org/10.1093/ptep/ptz106} {\path{doi:10.1093/ptep/ptz106}}.

\bibitem{Ablikim:2019hff}
M.~Ablikim, et~al., {Future Physics Programme of BESIII}, Chin. Phys. C 44~(4)
  (2020) 040001.
\newblock \href {http://arxiv.org/abs/1912.05983} {\path{arXiv:1912.05983}},
  \href {https://doi.org/10.1088/1674-1137/44/4/040001}
  {\path{doi:10.1088/1674-1137/44/4/040001}}.

\bibitem{Abada:2019lih}
A.~Abada, et~al., {FCC Physics Opportunities}: {Future Circular Collider
  Conceptual Design Report Volume 1}, Eur. Phys. J. C 79~(6) (2019) 474.
\newblock \href {https://doi.org/10.1140/epjc/s10052-019-6904-3}
  {\path{doi:10.1140/epjc/s10052-019-6904-3}}.

\bibitem{Abada:2019zxq}
A.~Abada, et~al., {FCC-ee: The Lepton Collider}: {Future Circular Collider
  Conceptual Design Report Volume 2}, Eur. Phys. J. ST 228~(2) (2019) 261--623.
\newblock \href {https://doi.org/10.1140/epjst/e2019-900045-4}
  {\path{doi:10.1140/epjst/e2019-900045-4}}.

\bibitem{Lai:2016uvj}
Y.-T. Lai, et~al., {Search for $D^{0}$ decays to invisible final states at
  Belle}, Phys. Rev. D 95~(1) (2017) 011102.
\newblock \href {http://arxiv.org/abs/1611.09455} {\path{arXiv:1611.09455}},
  \href {https://doi.org/10.1103/PhysRevD.95.011102}
  {\path{doi:10.1103/PhysRevD.95.011102}}.

\bibitem{Bause:2020obd}
R.~Bause, H.~Gisbert, M.~Golz, G.~Hiller, {Exploiting $CP$-asymmetries in rare
  charm decays}, Phys. Rev. D 101~(11) (2020) 115006.
\newblock \href {http://arxiv.org/abs/2004.01206} {\path{arXiv:2004.01206}},
  \href {https://doi.org/10.1103/PhysRevD.101.115006}
  {\path{doi:10.1103/PhysRevD.101.115006}}.

\bibitem{Golowich:2009ii}
E.~Golowich, J.~Hewett, S.~Pakvasa, A.~A. Petrov, {Relating D0-anti-D0 Mixing
  and D0 ---\ensuremath{>} l+ l- with New Physics}, Phys. Rev. D 79 (2009)
  114030.
\newblock \href {http://arxiv.org/abs/0903.2830} {\path{arXiv:0903.2830}},
  \href {https://doi.org/10.1103/PhysRevD.79.114030}
  {\path{doi:10.1103/PhysRevD.79.114030}}.

\bibitem{MartinCamalich:2020dfe}
J.~Martin~Camalich, M.~Pospelov, P.~N.~H. Vuong, R.~Ziegler, J.~Zupan, {Quark
  Flavor Phenomenology of the QCD Axion}, Phys. Rev. D 102~(1) (2020) 015023.
\newblock \href {http://arxiv.org/abs/2002.04623} {\path{arXiv:2002.04623}},
  \href {https://doi.org/10.1103/PhysRevD.102.015023}
  {\path{doi:10.1103/PhysRevD.102.015023}}.

\bibitem{Faisel:2020php}
G.~Faisel, J.-Y. Su, J.~Tandean, {Exploring charm decays with missing energy in
  leptoquark models} (12 2020).
\newblock \href {http://arxiv.org/abs/2012.15847} {\path{arXiv:2012.15847}}.

\bibitem{Dorsner:2016wpm}
I.~Dor\v{s}ner, S.~Fajfer, A.~Greljo, J.~Kamenik, N.~Ko\v{s}nik, {Physics of
  leptoquarks in precision experiments and at particle colliders}, Phys. Rept.
  641 (2016) 1--68.
\newblock \href {http://arxiv.org/abs/1603.04993} {\path{arXiv:1603.04993}},
  \href {https://doi.org/10.1016/j.physrep.2016.06.001}
  {\path{doi:10.1016/j.physrep.2016.06.001}}.

\bibitem{Bordone:2018nbg}
M.~Bordone, C.~Cornella, J.~Fuentes-Mart\'\i{}n, G.~Isidori, {Low-energy
  signatures of the $\mathrm{PS}^3$ model: from $B$-physics anomalies to LFV},
  JHEP 10 (2018) 148.
\newblock \href {http://arxiv.org/abs/1805.09328} {\path{arXiv:1805.09328}},
  \href {https://doi.org/10.1007/JHEP10(2018)148}
  {\path{doi:10.1007/JHEP10(2018)148}}.

\bibitem{DiLuzio:2017vat}
L.~Di~Luzio, A.~Greljo, M.~Nardecchia, {Gauge leptoquark as the origin of
  B-physics anomalies}, Phys. Rev. D 96~(11) (2017) 115011.
\newblock \href {http://arxiv.org/abs/1708.08450} {\path{arXiv:1708.08450}},
  \href {https://doi.org/10.1103/PhysRevD.96.115011}
  {\path{doi:10.1103/PhysRevD.96.115011}}.

\bibitem{Ruggiero}
G.~Ruggiero, {New Result on $K^+ \to \pi^+ \nu \bar\nu$ from the NA62
  Experiment}, KAON2019, Perugia, Italy, 10-13 September 2019.

\bibitem{Buras:2015qea}
A.~J. Buras, D.~Buttazzo, J.~Girrbach-Noe, R.~Knegjens, {$ {K}^{+}\to
  {\pi}^{+}\nu \overline{\nu} $ and $ {K}_L\to {\pi}^0\nu \overline{\nu} $ in
  the Standard Model: status and perspectives}, JHEP 11 (2015) 033.
\newblock \href {http://arxiv.org/abs/1503.02693} {\path{arXiv:1503.02693}},
  \href {https://doi.org/10.1007/JHEP11(2015)033}
  {\path{doi:10.1007/JHEP11(2015)033}}.

\bibitem{Angelescu:2020uug}
A.~Angelescu, D.~A. Faroughy, O.~Sumensari, {Lepton Flavor Violation and
  Dilepton Tails at the LHC}, Eur. Phys. J. C 80~(7) (2020) 641.
\newblock \href {http://arxiv.org/abs/2002.05684} {\path{arXiv:2002.05684}},
  \href {https://doi.org/10.1140/epjc/s10052-020-8210-5}
  {\path{doi:10.1140/epjc/s10052-020-8210-5}}.

\bibitem{Fuentes-Martin:2020lea}
J.~Fuentes-Martin, A.~Greljo, J.~Martin~Camalich, J.~D. Ruiz-Alvarez, {Charm
  physics confronts high-p$_{T}$ lepton tails}, JHEP 11 (2020) 080.
\newblock \href {http://arxiv.org/abs/2003.12421} {\path{arXiv:2003.12421}},
  \href {https://doi.org/10.1007/JHEP11(2020)080}
  {\path{doi:10.1007/JHEP11(2020)080}}.

\bibitem{Fajfer:2020tqf}
S.~Fajfer, D.~Susi\v{c}, {Coloured Scalar Mediated Nucleon Decays to Invisible
  Fermion} (10 2020).
\newblock \href {http://arxiv.org/abs/2010.08367} {\path{arXiv:2010.08367}}.

\bibitem{Fajfer:2015mia}
S.~Fajfer, N.~Ko\v{s}nik, {Prospects of discovering new physics in rare charm
  decays}, Eur. Phys. J. C 75~(12) (2015) 567.
\newblock \href {http://arxiv.org/abs/1510.00965} {\path{arXiv:1510.00965}},
  \href {https://doi.org/10.1140/epjc/s10052-015-3801-2}
  {\path{doi:10.1140/epjc/s10052-015-3801-2}}.

\bibitem{Carrasco:2015pra}
N.~Carrasco, P.~Dimopoulos, R.~Frezzotti, V.~Lubicz, G.~C. Rossi, S.~Simula,
  C.~Tarantino, {\ensuremath{\Delta}S=2 and \ensuremath{\Delta}C=2 bag
  parameters in the standard model and beyond from N$_f$=2+1+1 twisted-mass
  lattice QCD}, Phys. Rev. D 92~(3) (2015) 034516.
\newblock \href {http://arxiv.org/abs/1505.06639} {\path{arXiv:1505.06639}},
  \href {https://doi.org/10.1103/PhysRevD.92.034516}
  {\path{doi:10.1103/PhysRevD.92.034516}}.

\bibitem{Zyla:2020zbs}
P.~Zyla, et~al., {Review of Particle Physics}, PTEP 2020~(8) (2020) 083C01.
\newblock \href {https://doi.org/10.1093/ptep/ptaa104}
  {\path{doi:10.1093/ptep/ptaa104}}.

\bibitem{Amhis:2019ckw}
Y.~S. Amhis, et~al., {Averages of $b$-hadron, $c$-hadron, and $\tau$-lepton
  properties as of 2018} (9 2019).
\newblock \href {http://arxiv.org/abs/1909.12524} {\path{arXiv:1909.12524}}.

\bibitem{Fajfer:2008tm}
S.~Fajfer, N.~Kosnik, {Leptoquarks in FCNC charm decays}, Phys. Rev. D 79
  (2009) 017502.
\newblock \href {http://arxiv.org/abs/0810.4858} {\path{arXiv:0810.4858}},
  \href {https://doi.org/10.1103/PhysRevD.79.017502}
  {\path{doi:10.1103/PhysRevD.79.017502}}.

\bibitem{Burdman:2001tf}
G.~Burdman, E.~Golowich, J.~L. Hewett, S.~Pakvasa, {Rare charm decays in the
  standard model and beyond}, Phys. Rev. D 66 (2002) 014009.
\newblock \href {http://arxiv.org/abs/hep-ph/0112235}
  {\path{arXiv:hep-ph/0112235}}, \href
  {https://doi.org/10.1103/PhysRevD.66.014009}
  {\path{doi:10.1103/PhysRevD.66.014009}}.

\bibitem{Kamenik:2009kc}
J.~F. Kamenik, C.~Smith, {Tree-level contributions to the rare decays B+
  ---\ensuremath{>} pi+ nu anti-nu, B+ ---\ensuremath{>} K+ nu anti-nu, and B+
  ---\ensuremath{>} K*+ nu anti-nu in the Standard Model}, Phys. Lett. B 680
  (2009) 471--475.
\newblock \href {http://arxiv.org/abs/0908.1174} {\path{arXiv:0908.1174}},
  \href {https://doi.org/10.1016/j.physletb.2009.09.041}
  {\path{doi:10.1016/j.physletb.2009.09.041}}.

\bibitem{Fleischer:2019wlx}
R.~Fleischer, R.~Jaarsma, G.~Koole, {Testing Lepton Flavour Universality with
  (Semi)-Leptonic $D_{(s)}$ Decays}, Eur. Phys. J. C 80~(2) (2020) 153.
\newblock \href {http://arxiv.org/abs/1912.08641} {\path{arXiv:1912.08641}},
  \href {https://doi.org/10.1140/epjc/s10052-020-7702-7}
  {\path{doi:10.1140/epjc/s10052-020-7702-7}}.

\bibitem{Li:2020dpc}
G.~Li, T.~Wang, Y.~Jiang, J.-B. Zhang, G.-L. Wang, {Spin-$1/2$ invisible
  particles in heavy meson decays} (4 2020).
\newblock \href {http://arxiv.org/abs/2004.10942} {\path{arXiv:2004.10942}}.

\bibitem{DiLuzio:2019jyq}
L.~Di~Luzio, M.~Kirk, A.~Lenz, T.~Rauh, {$\Delta M_s$ theory precision
  confronts flavour anomalies}, JHEP 12 (2019) 009.
\newblock \href {http://arxiv.org/abs/1909.11087} {\path{arXiv:1909.11087}},
  \href {https://doi.org/10.1007/JHEP12(2019)009}
  {\path{doi:10.1007/JHEP12(2019)009}}.

\bibitem{Bazavov:2016nty}
A.~Bazavov, et~al., {$B^0_{(s)}$-mixing matrix elements from lattice QCD for
  the Standard Model and beyond}, Phys. Rev. D 93~(11) (2016) 113016.
\newblock \href {http://arxiv.org/abs/1602.03560} {\path{arXiv:1602.03560}},
  \href {https://doi.org/10.1103/PhysRevD.93.113016}
  {\path{doi:10.1103/PhysRevD.93.113016}}.

\bibitem{Dowdall:2019bea}
R.~Dowdall, C.~Davies, R.~Horgan, G.~Lepage, C.~Monahan, J.~Shigemitsu,
  M.~Wingate, {Neutral B-meson mixing from full lattice QCD at the physical
  point}, Phys. Rev. D 100~(9) (2019) 094508.
\newblock \href {http://arxiv.org/abs/1907.01025} {\path{arXiv:1907.01025}},
  \href {https://doi.org/10.1103/PhysRevD.100.094508}
  {\path{doi:10.1103/PhysRevD.100.094508}}.

\bibitem{Aoki:2019cca}
S.~Aoki, et~al., {FLAG Review 2019: Flavour Lattice Averaging Group (FLAG)},
  Eur. Phys. J. C 80~(2) (2020) 113.
\newblock \href {http://arxiv.org/abs/1902.08191} {\path{arXiv:1902.08191}},
  \href {https://doi.org/10.1140/epjc/s10052-019-7354-7}
  {\path{doi:10.1140/epjc/s10052-019-7354-7}}.

\bibitem{Lubicz:2017syv}
V.~Lubicz, L.~Riggio, G.~Salerno, S.~Simula, C.~Tarantino, {Scalar and vector
  form factors of $D \to \pi(K) \ell \nu$ decays with $N_f=2+1+1$ twisted
  fermions}, Phys. Rev. D 96~(5) (2017) 054514, [Erratum: Phys.Rev.D 99, 099902
  (2019), Erratum: Phys.Rev.D 100, 079901 (2019)].
\newblock \href {http://arxiv.org/abs/1706.03017} {\path{arXiv:1706.03017}},
  \href {https://doi.org/10.1103/PhysRevD.96.054514}
  {\path{doi:10.1103/PhysRevD.96.054514}}.

\bibitem{Becirevic:2020rzi}
D.~Be\v{c}irevi\'c, F.~Jaffredo, A.~Pe\~nuelas, O.~Sumensari, {New Physics
  effects in leptonic and semileptonic decays} (12 2020).
\newblock \href {http://arxiv.org/abs/2012.09872} {\path{arXiv:2012.09872}}.

\end{thebibliography}
\end{document}